\documentclass[showpacs,preprintnumbers,amsmath,amssymb]{revtex4}
\usepackage{mathrsfs, color}
\usepackage{stmaryrd}
\usepackage{cases}
\usepackage{amssymb}
\usepackage{amsmath, amsthm}
\usepackage{amsfonts}
\usepackage{graphicx}
\usepackage{amsmath,amstext,amsbsy,amssymb}
\renewcommand{\a}{\alpha}
\newcommand{\ra}{\rangle}

\def\Tr{\mathrm{Tr}}

\begin{document}
	\title[Uncertainty relations for metric adjusted skew information and Cauchy-Schwarz inequality]
	{Uncertainty relations for metric adjusted skew information and Cauchy-Schwarz inequality}
	
	\author{Xiaoli~Hu}
	\email{xiaolihumath@jhun.edu.cn}
	\affiliation{School of Artificial Intelligence, Jianghan University, Wuhan, Hubei 430056, China}
	
	\author{Naihuan Jing}
	\email{jing@ncsu.edu}
	\affiliation{Department of Mathematics, North Carolina State University, Raleigh, NC 27695, USA}
	\date{\today}
	\begin{abstract}
Skew information is a pivotal concept in quantum information, quantum measurement, and quantum metrology. Further studies have lead to the 
uncertainty relations grounded in metric-adjusted skew information. In this work, we present an in-depth investigation using the methodologies of sampling coordinates of observables and convex functions to refine the uncertainty relations in both the product form of two observables and summation form of multiple observables.
% with connections with group representations.
%also establish stronger lower bounds in
	\end{abstract}
	
	\keywords{uncertainty relations, skew information, Cauchy-Schwarz inequality}
	\maketitle
	%%%%%%%%%%%%%%%%%%%%%%%%%%%%%%%%%%%%%%%%%%%%%%%%%%%%
	
\section{Introduction}
Quantum mechanics is a fundamental theory that describes the behavior of matter and energy at the atomic and subatomic scale \cite{DP}. One of its most distinctive features is the Heisenberg uncertainty principle, which states that certain pairs of physical properties, such as position and momentum, cannot be simultaneously measured with arbitrary precision \cite{WH, KEH}.
The generalization of Heisenberg's uncertainty principle to encompass any two non-commuting observables was accomplished by Robertson \cite{HPR} and later refined by Schr\"odinger \cite{Schr}. Although uncertainty relations play a crucial role in the study of quantum systems, there are other types of uncertainty in quantum systems that are not captured by the standard uncertainty relations, for instance the uncertainty of skew information. This concept has important applications in quantum information processing, such as in the design of more accurate and efficient quantum measurements.

Skew information was first introduced by Wigner and Yanase in 1963 as a measure of the information content of quantum states in the presence of conserved quantities \cite{Wigner2}. For a composite system, such as a measuring device consisting of a system of interest and
a pointer, Wigner demonstrated that it is impossible to perform an ideal measurement of an observable that does not
commute with a conserved quantity in the presence of the conservation law \cite{Wigner1}. Luckily, performing an approximate
measurement of the observable is still feasible, and the error diminishes as the size of the system grows \cite{YMM1,YMM2}. Such a trade-off between the measuring accuracy and the size of the measuring apparatus is known as Wigner-Araki-Yanase
theorem in literature. Further remarkable studies were later presented in the references \cite{OM1, MS, KS, OM2,OM3, FBR}.

In recent years, there has been significant interest in developing new techniques for quantifying and understanding skew information in quantum systems.
One important direction of research has been to take into account the geometry of the underlying quantum state space. This has led to the development of the metric-adjusted skew information, which is a measure of skew information that takes into account the curvature of the quantum state space \cite{MH06}. Uncertainty relations based on metric-adjusted skew information can establish fundamental limits on quantum measurement accuracy and information processing speed. Additionally, skew information can help understand quantum correlations and classical behavior in quantum systems.
Recent developments in uncertainty relations include the introduction of a new type of uncertainty relation using metric-adjusted skew information \cite{SL}, which extended the existing Wigner-Yanase-Dyson (WYD) skew information and demonstrated its practical importance.
The results indicate that the skew information provides a new way to quantify Bohr’s complementary principle and Heisenberg’s uncertainty principle. The Wigner-Yanase-Dyson metric adjusted
skew information was introduced as a generalization of the Wigner-Yanase skew information in \cite{Hansen}. In \cite{Cai},
it was demonstrated that the sum uncertainty relation for the Wigner-Yanase skew information, introduced in \cite{Chen},
also holds for arbitrary metric-adjusted skew information, and the related uncertainty relations for quantum channels were given there. An equivalent formulation for arbitrary two observables and two channels was also presented in \cite{Ren}, and a generalization to multi-observables was given in \cite{CWL, ZF}. Recently, product and summation forms of the uncertainty relation were proposed in \cite{Ma} for the WYD-metric adjusted skew information based on two different refinements of the Cauchy-Schwarz inequality.

In this work, we introduce novel and improved uncertainty relations in both product and summation forms for two and $N$ quantum observables, respectively. Our approach involves utilizing carefully chosen convex functions to enhance the Cauchy-Schwarz inequality and measuring the observables using sampled measurement coordinates. This work presents new uncertainty relations for the WYD-metric adjusted skew information, which provide a tighter bound on the amount of uncertainty in a quantum system. Our motivation stems from an information-theoretic perspective, where several information measures are crucial to the theory's structural properties.
Overall, the study of uncertainty and skew information in quantum systems is an important and active area of research, with many potential applications in quantum information processing and other areas of science and technology.

 \section{Preliminaries}
		Recall that the Wigner-Yanase skew information \cite{Wigner2} is a measure of non-commutativity between the square root of the state $\rho$ and the conserved observable $A$: 
	\begin{equation}
		\begin{split}
			I_{\rho}(A)=-\frac{1}{2}\Tr([A,\sqrt{\rho}])^2
			=\Tr(\rho A^2)-\Tr(\sqrt{\rho}A\sqrt{\rho}A),
		\end{split}
	\end{equation}
	where Tr means the trace operation, $[A, \sqrt{\rho} ]$ is the commutator of $A$ and $\sqrt{\rho}$, so 
it vanishes when $\rho$ commutes with $A$. $I_{\rho}$ is clearly homogeneous in $\rho$.
	When $\rho=|\psi\rangle \langle \psi|$ is a pure state, $I_{\rho}(A)=\Tr(\rho A^2)-(\Tr(\rho A))^2=\Delta_{\rho}(A)$ is the variance of the observable $A$ with respect to $\rho$. This means that the Wiger-Yanase skew information generalizes the variance. Furthermore, when $\rho$ is a mixed state, we have $I_{\rho}(A)>\Delta_{\rho}(A)$ in general. The Heisenberg uncertainty principle in terms of the skew information states that \cite{SL}:
	\begin{equation}
		I_{\rho}(A)I_{\rho}(B)\geq \frac{1}{4}|\Tr(\rho[A,B])|^2.
	\end{equation}
	
	The Fisher-Rao metric for classical information is the unique contracting Riemannian metric under the Markov morphism \cite{Hansen}. The symmetric monotone metric for quantum case is constructed through the so-called Morozova-Chentsov functions. Such a function $f:(0,+\infty)\to \mathbb{R}$ is a positive operator monotonic function (i.e. $0<f(A)<f(B)$ holds when $0<A<B$ ) such that $f$ is symmetric (i.e. $f(x) = xf(x^{-1})$), and normalized (i.e. $f(1)=1$).
	The Morozova-Chentsov function corresponding to a positive operator-monotone and symmetric function $f$ is defined by
	\begin{equation}
		c(x,y)=\frac{1}{yf(xy^{-1})},\ \   x, y>0.
	\end{equation}
	
	 According to Morozova, Chentsov and Petz,
	the monotone metric on the state space of a quantum system is given by (cf. \cite{Petz})
	\begin{equation}
		K^c_{\rho}(A,B)=\Tr(A^{\dag}c(L_{\rho},R_{\rho})B),
	\end{equation}
where $L_{\rho}(A)=\rho A$ and $R_{\rho}(A)=A\rho$ be left and right multiplication operators, then $[\rho, A]=\rho A-A\rho=(L_{\rho}-R_{\rho})A$.
	This metric form depends on the choice of the Morozova-Chentsov function, so it is not unique. The metric constant is denoted by $m(c)=\lim_{t\to 0}c(t,1)^{-1}=\lim_{t\to 0}f(t)$. A symmetric monotone metric is called regular if $m(c)>0$. Because $f(t)$ is a positive operator monotone function on $[0, \infty)$, we have $m(c)=f(0)$ for a regular metric.
	Let $c$ be the Morozova-Chentsov function of a regular metric. Then the so-called metric adjusted skew information $I_{\rho}^c(A)$ for observable $A$ and quantum state $\rho$ is defined by \cite{KS, Hansen}:
	\begin{equation}
		I_{\rho}^c(A)=\frac{m(c)}{2} K_{\rho}^c(\mathbf{i}[\rho,A],\mathbf{i}[\rho,A])=\frac{m(c)}{2} \Tr(\mathbf{i}[\rho,A]c(L_{\rho},R_{\rho})\mathbf{i}[\rho,A]),
	\end{equation}
	where $\mathbf{i}^2=-1$. If we consider the following positive operator monotone and symmetric function
	\begin{equation}
		f_{p}(x)=p(1-p)\frac{(x-1)^2}{(x^{p}-1)(x^{1-p}-1)}, \qquad 0<p<1,
	\end{equation}
	then the corresponding Morozova-Chentsov function is of the form
	\begin{equation}\label{e:c}
		c^{WYD}(x,y)=\frac{1}{p(1-p)}\frac{(x^p-y^p)(x^{1-p}-y^{1-p})}{(x-y)^2}, \ \ 0<p<1.
	\end{equation}
	Therefore $m(c^{WYD})=\lim_{t\to 0}c^{WYD}(t,1)=p(1-p)$ and the WYD-metric adjusted skew information is given by
	\begin{equation}\label{e:WYD}
		I_{\rho}^{c^{WYD}}(A)=-\frac{1}{2}\Tr([\rho^{p},A][\rho^{1-p},A]), \ \  0<p<1.
	\end{equation}
	Equation (\ref{e:WYD}) is the Dyson generalization of the Wigner-Yanase skew information when $p=\frac{1}{2}$. The famous Wigner-Yanase-Dyson (WYD) conjecture on the convexity of $I_{\rho}^{c^{WYD}}(A)$ was proved by Lieb \cite{Lieb}.
	
	When $c$ is a regular Morozova-Chentsov function, the metric-adjusted correlation of observables $A$ and $B$ is given by
	\begin{equation}
		\mathrm{Corr}_{\rho}^c(A,B)=\frac{m(c)}{2}K_{\rho}^c(\mathbf{i}[\rho,A],\mathbf{i}[\rho,B]).
	\end{equation}
	
The metric adjusted correlation is a measure of statistical dependence between two observables in quantum mechanics that takes into account their non-commutativity. It's a useful tool for detecting entanglement and studying quantum phase transitions. However, it's not a real-valued function and can take complex values, so it can't directly provide positive bounds. We present a novel approach that addresses this issue by utilizing coordinate sampling of observables and a carefully chosen convex function, in combination with the Cauchy-Schwarz inequality, to derive robust uncertainty relations for the WYD-metric adjusted correlation. Our method yields stronger bounds than several previous uncertainty relations.
	
The paper is organized as follows. In Section II, we provide a review of two recent uncertainty relations that offer a ``fine-grained'' analysis of the product-form uncertainty relation for two observables using the Cauchy-Schwarz inequality. We then introduce our technique of coordinate sampling and convex function selection to further enhance the lower bound. In Section III, we present a set of practical and straightforward uncertainty relations for the sum-form skew information based on the fundamental inequality and well-chosen convex functions. We conclude in Section IV.
	
	%%%%%%%%%%%%%%%%%%%%%%%%%%%%%%%%%%%%%%%%%%%%%%%%%%%%
	
	\section{Improved uncertainty relations for product-form skew information}
	Consider a Hilbert space $H$ of dimensional $d$ over the complex field $\mathbb{C}$. Let $M_d(\mathbb{C})$ denote by the set of complex matrices and $D_d(\mathbb{C})$ the set of complex density matrices on $H$. We use  $\langle X, Y \rangle= \Tr(X^{\dag}Y)$ to denote the Hilbert-Schmidt scalar product, where $X, Y\in M_d(\mathbb{C})$. The computational basis of $H$ is given by $\{|i\rangle, i=1,2,\cdots,d\}$, and $\{E_{ij}=|i\rangle\langle j|, i,j=1,2,\cdots,d\}$ forms a complete set of local orthogonal matrices. Specifically, 	$\langle E_{ij},E_{kl}\rangle=\Tr(E_{ji}E_{kl})=\delta_{ik}\delta_{jl}$, and they form an orthogonal basis for all observables in $M_d(\mathbb{C})$. Given a matrix $M=[m_{ij}]\in M_d(\mathbb{C})$, we define  $\vec{M}=(m_{11},m_{12},\cdots,m_{1d},m_{21},\cdots,m_{dd})^{T}$ as its $d^2$-dimensional coordinate vector with respect to the
	standard basis $\{E_{ij}\}$.
Let $\overrightarrow{\mathbf{E}}=(E_{11}, E_{12},\cdots, E_{1d}, E_{21},\cdots,E_{dd})^{T}$, then any observable 
$A$ can be expressed as a linear combination of these basis observables, i.e.
	\begin{equation}
		A=\sum_{i,j=1}^da_{ij}E_{ij}=\langle \vec{A}, \overrightarrow{\mathbf{E}}\rangle=\vec{A}^{ \dag}\cdot\overrightarrow{\mathbf{E}}\ 
	\end{equation}
	with $a_{ij}=\langle E_{ij}, A\rangle=\Tr(E_{ji}A)$.

	%where $\vec{A}$ is the vector of $A$.
	The metric adjusted skew information of $A$ is given by
	\begin{equation}\label{Gamma}
		\begin{split}
			I_{\rho}^c(A)&=\frac{m(c)}{2}K_{\rho}^c(\mathbf{i}[\rho,A],\mathbf{i}[\rho,A])\\
			&=\frac{m(c)}{2}\sum_{i,j=1}^d\sum_{k,l=1}^d a_{ij}^*a_{kl}K_{\rho}^c(\mathbf{i}[\rho,E_{ij}],\mathbf{i}[\rho,E_{kl}])\\
			%&=\frac{m(c)}{2}\sum_{ij}\sum_{k,l} a_{ij}a_{kl}Tr(\mathbf{i}[\rho,E_{ij}]^{\dag}c(L_{\rho}, R_{\rho})\mathbf{i}[\rho,E_{kl}])\\
			&=\vec{A}^{\dag}\cdot\Gamma\cdot\vec{A},
		\end{split}
	\end{equation}
	where $\Gamma=(\Gamma_{i_j,k_l}) $ and $\Gamma_{i_j,k_l}=\frac{m(c)}{2}K_{\rho}^c(\mathbf{i}[\rho,E_{ij}],\mathbf{i}[\rho,E_{kl}])=\mathrm{Corr}_{\rho}^c(E_{ij},E_{kl})$ with $i_j=d(i-1)+j, k_l=d(k-1)+l$. In fact, under the Wigner-Yanase-Dyson-metric we have
	\begin{equation}\label{e:C1}
		\begin{split}
			\Gamma_{i_j,k_l}%&=\frac{m(c)}{2}K_{\rho}^c(\mathbf{i}[\rho,E_{ij}],\mathbf{i}[\rho,E_{kl}])\\
			&=\frac{m(c^{WYD})}{2}\Tr((\mathbf{i}[\rho,E_{ij}])^{\dag}c^{WYD}(L_{\rho}, R_{\rho})\mathbf{i}[\rho,E_{kl}])\\
			&=-\frac{1}{2}\Tr[(L_{\rho}-R_{\rho})E_{ji}\frac{(L_{\rho}^p-R_{\rho}^p)(L_{\rho}^{1-p}-R_{\rho}^{1-p})}{(L_{\rho}-R_{\rho})^2} (L_{\rho}-R_{\rho})E_{kl})]\\
			&=-\frac{1}{2}\Tr([\rho^p,E^{\dag}_{ij}][\rho^{1-p},E_{kl}])\\
			&=\frac{1}{2}\langle[\rho^p,E_{ij}], [\rho^{1-p},E_{kl}]\rangle.
		\end{split}
	\end{equation}
	It follows from \eqref{Gamma} that $\Gamma$ is semi-positive, so there is a matrix $C$ such that $\Gamma=C^{\dag}C$. Then $I_{\rho}^c(A)=\vec{A}^{\dag}C^{\dag}C\vec{A}=(C\vec{A})^{\dag}(C\vec{A})=|\vec{\a}|^2$,
	where $\vec{\a}=C\vec{A}=(\a_1,\cdots,\a_{d^2})^{\dag}$. Let $\vec{B}$ be the coordinate vector of the quantum observable $B$. Similarly $I_{\rho}^c(B)=|\vec{\beta}|^2$ with $\vec{\beta}=C\vec{B}=(\beta_1,\cdots,\beta_{d^2})^{\dag}$. Therefore,
the product of the Wigner-Yanase-Dyson-metric adjusted skew information for observables $A$ and $B$ can be written as
	\begin{equation}\label{e:re}
		\begin{split}
			I_{\rho}^{c^{WYD}}(A)I_{\rho}^{c^{WYD}}(B)&=|\vec{\a}|^2|\vec{\beta}|^2=\sum_{i,j=1}^{d^2}|\a_i|^2|\beta_j|^2\geq |\sum_{i=1}^{d^2} \a_i^*\beta_i|^2=|(\vec{\a},\vec{\beta})|^2=|\mathrm{Corr}_{\rho}^{c^{WYD}}(A,B)|^2,
		\end{split}
	\end{equation}
which is the usual form of the adjusted metric skew information based uncertainty relation.
	
	Consider an $n$-dimensional real vector space $\mathbb{R}^n$ equipped with the classical inner product $( \ , \ )$. Let $\vec{v}=(v_1,\cdots,v_n)^T\in \mathbb{R}^n$ be a vector, and let $\vec{v}_k=(v_1,\cdots,v_k,0,\cdots,0)^T$  denote a partial vector of $\vec{v}$ with the first $k$ components. We define the complement vector $\vec{v}^c_k=\vec{v}-\vec{v}_k$.
	The symmetric group  $\mathfrak{S}_n$ acts on $\mathbb R^n$ by permutation, i.e. for $\sigma\in \mathfrak{S}_n$, the action is given by $\sigma(\vec{v})=(v_{\sigma(1)},\cdots, v_{\sigma(n)})^T$. In particular, for any $\sigma\in \mathfrak{S}_n$ and for any $k\leq n$, we have
	$\sigma(\vec{v}_k)=(v_{\sigma(1)},\cdots, v_{\sigma(k)},0,\cdots,0)^T$, then
	$\sigma(\vec{v}_k^c)=\sigma(v)-\sigma(\vec{v}_k)$.

	In the sequel, we set $n=d^2$, where $\dim(H)=d$. We define $x_i=|\a_i|$ and $y_j=|\beta_j|$  for $i,j=1,\cdots, n$ in (\ref{e:re}), so that $I^{c^{WYD}}_{\rho}(A)I^{c^{WYD}}_{\rho}(B)=\sum_{i, j=1}^nx_i^2y_j^2$. The $n^2$ elements $x_i^2y_j^2$ can be stored in
	a real matrix $M_n=[x_i^2y_j^2]_{1\leq i, j\leq n}$. Let $M_k$ be the $k$-th sequential principal matrix of $M_n$. More generally, 
	let $M_I=[x_i^2y_j^2]_{i, j\in I}$ for $I\subset\{1,2 \ldots, n\}$. The norm of $M_k$ is denoted by
	$|M_k|_{\infty}=\sum _{i,j=1}^kx_i^2y_j^2$.
	 In particular,
	\begin{equation}
		|M_n|_{\infty}=\sum_{i,j=1}^n x_i^2y_j^2=I^{c^{WYD}}_{\rho}(A)I^{c^{WYD}}_{\rho}(B).
	\end{equation}
	For partial vectors $\vec{x_k}=(x_1,\cdots, x_k,0,\cdots,0)^T$, $\vec{y_k}=(y_1,\cdots, y_k,0,\cdots,0)^T\in \mathbb{R}^n$, let $f_{CS}$ be the function defined by $f_{CS}(M_k)=(\vec{x}_k,\vec{y}_k)^2$,
	and similarly defined for a general principal matrix. By the Cauchy-Schwarz inequality, we have 	\begin{equation}|M_k|_{\infty}=\sum _{i,j=1}^kx_i^2y_j^2=(\vec{x}_k,\vec{x}_k)(\vec{y}_k,\vec{y}_k)\geq (\vec{x}_k,\vec{y}_k)^2=f_{CS}(M_k).
	\end{equation} The quantity proposed by Xiao et al. in \cite{XJ3} and Yu et al. in \cite{Yu} is given by this expression for  $k\leq n$, i.e.
	\begin{equation}\label{e:Ik}
		\begin{split}
			I_k&=|M_n|_{\infty}-|M_k|_{\infty} +f_{CS}(M_k)=\sum_{i=1}^nx_i^2y_i^2+\sum_{1\leq i<j\leq n, k<j}(x_i^2y_j^2+x_j^2y_i^2)
			+\sum_{1 \leq i<j\leq k}2x_iy_ix_jy_j.
		\end{split}
	\end{equation}
	We will mainly use $I_2$, $ I_3$ and $I_4$ in our later examples: 
	\begin{equation}\label{e:I2}
		\begin{split}
			I_2=&\sum_{i,j=1}^nx_i^2y_j^2-x_1^2y_2^2-x_2^2y_1^2+2x_1x_2y_1y_2;\\
			I_3=&\sum_{i,j=1}^nx_i^2y_j^2-x_1^2y_2^2-x_2^2y_1^2-x_1^2y_3^2-x_3^2y_1^2-x_2^2y_3^2-x_3^2y_2^2
			+2x_1x_2y_1y_2+2x_1x_3y_1y_3+2x_2x_3y_2y_3;\\
			I_4=&\sum_{i,j=1}^nx_i^2y_j^2-x_1^2(y_2^2+y_3^2+y_4^2)-x_2^2(y_3^2+y_4^2)-x_3^2y_4^2-y_1^2(x_2^2+x_3^2+x_4^2)-y_2^2(x_3^2+x_4^2)-y_3^2x_4^2\\
			&+2x_1x_2y_1y_2+2x_1x_3y_1y_3+2x_1x_4y_1y_4+2x_2x_3y_2y_3+2x_2x_4y_2y_4+2x_3x_4y_3y_4.\\
		\end{split}
	\end{equation}

	See FIG.1 for the meaning of $M_k$ and $I_k$.
	\begin{figure}[!ht]
		%\centering
		\includegraphics[width=8cm,height=4.5cm]{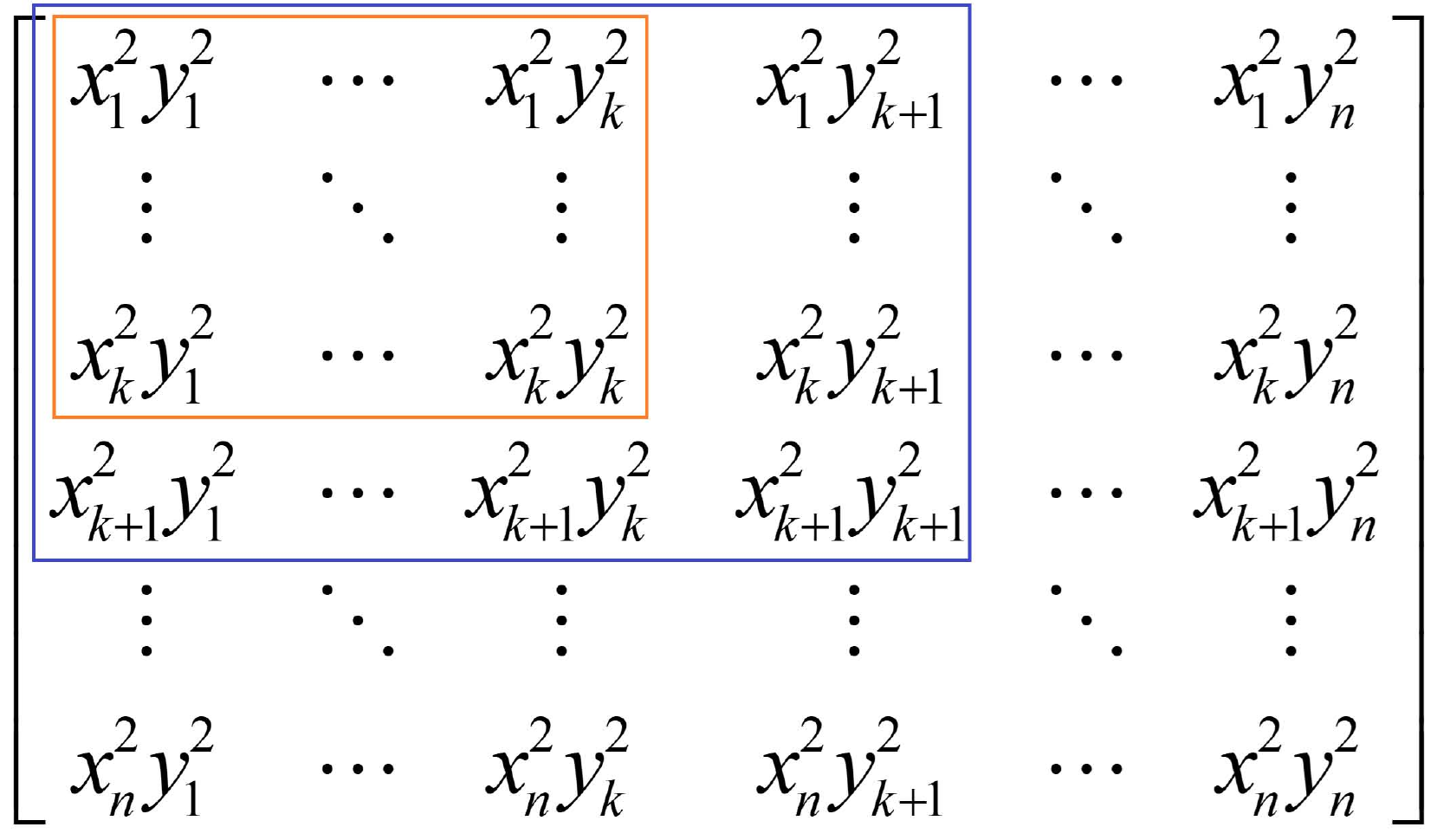}
		\caption{{\footnotesize\textbf{The matrix $M_k$ and quantity $I_k$.} The part framed in orange and blue are the sequential principal matrices $M_k$ and $M_{k+1}$ respectively. $I_k$ is equal to $f_{CS}(M_k)$ plus the elements outside of the $k$-th sequential principal matrix $M_k$.}} \label{gra1}
	\end{figure}
	In practice, the storage of $n^2$ elements $x_i^2y_j^2 ~(i,j=1,\cdots,n)$ in a matrix is random. The randomness can be generated by the action of  the symmetry group $\mathfrak{S}_n$. For $\sigma, \tau\in \mathfrak{S}_n$, we define $M_n^{\sigma,\tau}=(\sigma, \tau)M_n=(m_{\sigma(i),\tau(j)})_{n\times n}$, then $ |M^{\sigma, \tau}_n|_{\infty}=|M_n|_{\infty}$ and
	\begin{equation}\label{eq:IIk}
		\begin{split}
			(\sigma,\tau)\cdot I_k&=|M^{\sigma, \tau}_n|_{\infty}-|M^{\sigma, \tau}_k|_{\infty} + f_{CS}(M_k^{\sigma, \tau})\\
			&=\sum_{i=1}^nx_{\sigma(i)}^2y_{\tau(i)}^2+\sum_{1\leq i<j\leq n,k<j}(x_{\sigma(i)}^2y_{\tau(j)}^2+x_{\sigma(j)}^2y_{\tau(i)}^2)
			+\sum_{1\leq i<j\leq k}2x_{\sigma(i)}y_{\tau(j)}x_{\sigma(j)}y_{\tau(i)}.
		\end{split}
	\end{equation}
	
Let $T_1=\{\vec{t}=(t_1,\cdots,t_{d^2})|0\leq t_i \leq 1, \sum_{i=1}^{d^2} t_i=1\}$ be the principal region of the unit ball.  For $\vec{t} \in T_1$,  we denote by $I^{\vec{t}}=\sum_{i=1}^{d^2}t_iI_i$ the convex combination of $I_1,I_2,\cdots,I_{d^2}$, then we have the following result.
	\medskip
	
	\emph{Theorem 1.}\label{t:I} For any two observables $A$ and $B$ with a state $\rho$ on a $d$-dimensional Hilbert space $H$, the product of the Wigner-Yanase-Dyson-metric adjusted skew information of $A$ and $B$ obeys the following uncertainty relations:
	\begin{equation}
		\begin{split}
			(i)~ I^{c^{WYD}}_{\rho}(A)I^{c^{WYD}}_{\rho}(B)&\geq \max_{\vec{t} \in T_1}\{I^{\vec{t}} \}\geq|\mathrm{Corr}^{c^{WYD}}_{\rho}(A,B)|^2;\\
			(ii)~I^{c^{WYD}}_{\rho}(A)I^{c^{WYD}}_{\rho}(B)&\geq qI_1+(1-q)\max_{\sigma,\tau\in\mathfrak{S}_{d^2}}\{(\sigma,\tau)I_k\}\geq |\mathrm{Corr}^{c^{WYD}}_{\rho}(A,B)|^2.
		\end{split}
	\end{equation}
	where $\vec{t}\in T_1$ and $0\leq q\leq 1$.
%\emph{Proof:}
\begin{proof} 	From FIG.1, we can see that $I_k- I_{k-1}=-(\sum_{i=1}^{k-1} x_{i}y_k+y_ix_k)^2\leq 0$.
	Then the following descending sequence of inequalities holds:	
	\begin{equation}\label{e:re21}
		 I^{c^{WYD}}_{\rho}(A)I^{c^{WYD}}_{\rho}(B)=I_1\geq I_2\geq\cdots \geq I_{d^2}=|\mathrm{Corr}^{c^{WYD}}_{\rho}(A,B)|^2.
	\end{equation}
	 Moreover, $I_1\geq \max_{\sigma,\tau\in \mathfrak{S}_{d^2}}\{(\sigma,\tau)I_k\}$ for any permutation pairs $(\sigma,\tau) \in \mathfrak{S}_{d^2}\times \mathfrak{S}_{d^2}$. Then we have the following 
	\begin{equation}\label{e:re22}
		\begin{split}
		 I^{c^{WYD}}_{\rho}(A)I^{c^{WYD}}_{\rho}(B)&\geq\max_{\sigma,\tau\in\mathfrak{S}_{d^2}}\{(\sigma,\tau)I_k\}\geq |\mathrm{Corr}^{c^{WYD}}_{\rho}(A,B)|^2.
		\end{split}
	\end{equation}
%he uncertainty relations \eqref{e:re21} and  \eqref{e:re22} can see in \cite{Ma}. 
The theorem follows from \eqref{e:re21}, \eqref{e:re22} and the properties of the convex functions. \end{proof}
	
	Next, we introduce the second uncertainty relation for the WYD-metric adjusted skew information.
	%Let $g$ be a function such that $g(x_i^2y_j^2+x_j^2y_i^2)=2x_iy_jx_jy_i$.
	\begin{figure}[!ht]
		%\centering
		\includegraphics[width=8cm,height=4.5cm]{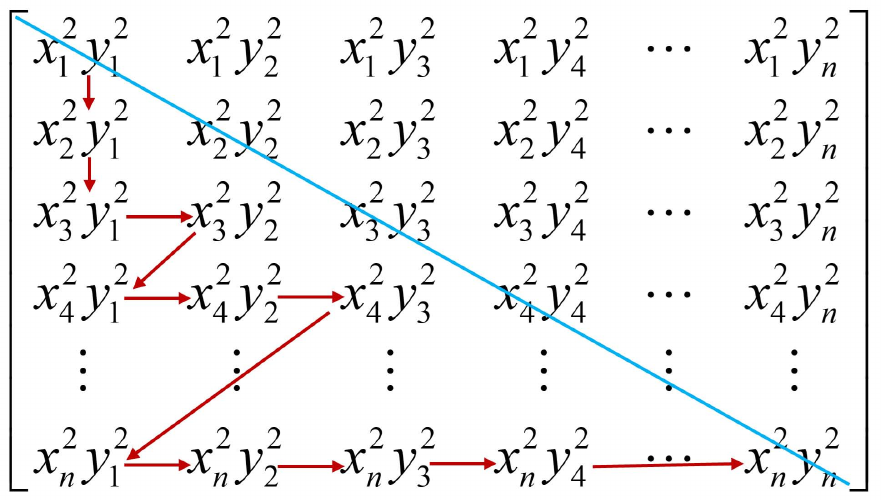}
		\caption{{\footnotesize\textbf{Explanation of $S_{pq}$.}  The Cauchy-Schwarz inequality is used sequentially from top to bottom in the direction of the arrow. Set $|M_n|_{\infty}=S_{10}=\sum_{i,j=1}^n x_i^2y_j^2$ as the initial value, the second is  $S_{21}=S_{10}-\sum_{i,j\in\{2,1\}}x_i^2y_j^2+f_{CS}(M_{\{2,1\}})$; the third is $S_{31}=S_{21}-\sum_{i,j\in\{3,1\}}x_i^2y_j^2+f_{CS}(M_{\{3,1\}})$; the fourth is $S_{32}=S_{31}-\sum_{i,j\in\{3,2\}}x_i^2y_j^2+f_{CS}(M_{\{3,2\}})$; the fifth is $S_{41}=S_{32}-\sum_{i,j\in\{4,1\}}x_i^2y_j^2+f_{CS}(M_{\{4,1\}})$, and the last term
			$S_{n,n-1}=S_{n,n-2}-\sum_{i,j\in\{n,n-1\}}x_i^2y_j^2+f_{CS}(M_{\{n, n-1\}})$.}}\label{gra2}
	\end{figure}
	 Recall the quantities proposed by Li et al. in \cite{JSL}:
	\begin{equation}\label{e:Spq}
		\begin{split}
			S_{pq}&=S_{p,q-1}-\sum_{i,j\in\{p,q\}}x^2_{i}y^2_{ j}+f_{CS}(M_{i,j\in\{p,q\}}),\\
			S_{p+1,1}&=S_{p,p-1}-\sum_{i,j\in\{p+1,1\}}x^2_iy^2_{j}+f_{CS}(M_{i,j\in\{p+1,1\}}),
		\end{split}
	\end{equation}
	where $0\leq q\leq p-1$ (see FIG. 2 for an explanation).
	%{\color{red} FIG.2 is the explanation of $S_{pq}$.
	For example,
	\begin{equation}\label{e:S21}
		\begin{split}
			S_{21}&=\sum_{i,j=1}^nx_i^2y_j^2-x_1^2y_2^2-x_2^2y_1^2+2x_1x_2y_1y_2;\\
			S_{31}&=\sum_{i,j=1}^nx_i^2y_j^2-x_1^2y_2^2-x_2^2y_1^2-x_1^2y_3^2-x_3^2y_1^2+2x_1x_2y_1y_2+2x_1x_3y_1y_3.
		\end{split}
	\end{equation}
	
	By the randomness of the entries in $M_n$ along the horizontal direction in FIG.2, we have
	\begin{equation}
		(\sigma,\tau)\cdot S_{pq}=(\sigma,\tau)\cdot S_{p,q-1}-\sum_{i,j\in\{\sigma(p),\tau (q-1)\}}x_{i}y_{ j}+f_{CS}(M_{i,j\in\{\sigma(p),\tau (q-1)\}}),
	\end{equation}
	along the left slash direction in FIG.2 we also have
	\begin{equation}
		(\sigma,\tau)\cdot S_{p+1,1}=(\sigma,\tau)\cdot S_{p,p-1}-\sum_{i,j\in\{\sigma(p+1),\tau(1)\}}x_iy_{j}+f_{CS}(M_{\{\sigma(p+1),\tau(1)\}}).
	\end{equation}

	Let $S^{\vec{t}}=\sum_{1\leq j<i\leq n}t_{ij}S_{ij}$ for $\vec{t}=(t_{10}, t_{21}, t_{31}, t_{32}, t_{41}, t_{42}\cdots)$, then we have the following result.
	\medskip
	
	\emph{Theorem 2.} \label{t:S} For any two observables $A$ and $B$ associated to a state $\rho$ on a $d$-dimensional Hilbert space $H$, the following uncertainty relations hold for the Wigner-Yanase-Dyson-metric adjusted  skew information:
	\begin{equation}
		\begin{split}
			(i)~I^{c^{WYD}}_{\rho}(A)I^{c^{WYD}}_{\rho}(B)&\geq \max_{t\in T_2}\{S^{t}\}\geq|\mathrm{Corr}^{c^{WYD}}_{\rho}(A,B)|^2,\\
			(ii)~I^{c^{WYD}}_{\rho}(A)I^{c^{WYD}}_{\rho}(B)&\geq qS_{10}+(1-q)\max_{\sigma,\tau\in\mathfrak{S}_{d^2}}\{
			(\sigma,\tau)S_{pq}\}\geq|\mathrm{Corr}^{c^{WYD}}_{\rho}(A,B)|^2,
		\end{split}
	\end{equation}
	where $ T_2=\{\vec{t}=(t_{10},t_{21},\cdots,t_{pq},\cdots)|0 \leq t_{ij}\leq 1,\sum_{1\leq j<i\leq d^2}t_{ij}=1\}$.
	\begin{proof} By \eqref{e:Spq} and the Cauchy-Schwarz inequality, we have the following sequence of inequalities (cf. \cite{JSL}):
		\begin{equation}\label{e:re31}
			I^{c^{WYD}}_{\rho}(A)I^{c^{WYD}}_{\rho}(B)=S_{10}\geq S_{21}\geq \cdots \geq S_{d^2,d^2-1}\geq|\mathrm{Corr}^{c^{WYD}}_{\rho}(A,B)|^2,
		\end{equation}
	where the subscript $(k,l)$ of $S_{k,l}$ satisfies
		$k>l$. 
	
 As $S_{pq}\leq \max_{\sigma,\tau\in\mathfrak{S}_{d^2}}\{(\sigma,\tau)\cdot S_{pq}\} $, we get the following inequalities: 
		\begin{equation}\label{e:re32}
	 I^{c^{WYD}}_{\rho}(A)I^{c^{WYD}}_{\rho}(B)\geq\max_{\sigma,\tau\in\mathfrak{S}_{d^2}}\{
			(\sigma,\tau)S_{pq}\}\geq |\mathrm{Corr}^{c^{WYD}}_{\rho}(A,B)|^2.\\
		\end{equation}
	The results of the theorem follow easily from \eqref{e:re31}, \eqref{e:re32} and the properties of convex functions. \end{proof}
		
		Note that Theorem 2 is a refinement of Theorem 1, as $S_{k,k-1}=I_k$.
	
We will now derive a third uncertainty relation for the WYD-metric adjusted skew information. The quantities obtained by sampling observable coordinates in \cite{HJ} are given as follows:
	\begin{equation}\label{e:Kk1}
		K_k=(|\vec{x}_k||\vec{y}_k|+|\vec{x}^c_k||\vec{y}^c_k|)^2,\ \ k=1,\cdots,n.
	\end{equation}
	Explicitly %It is clear that
	\begin{equation}\label{e:K2}
		\begin{split}
			K_n&=|\vec{x}_n|^2|\vec{y}_n|^2=I^{c^{WYD}}_{\rho}(A)I^{c^{WYD}}_{\rho}(B);\\
			K_1&=\big(\sqrt{x_1^2y_1^2}+\sqrt{(x_2^2+\cdots+x_n^2)(y_2^2+\cdots+y_n^2)} \big)^2;\\
			K_2&=\big(\sqrt{(x_1^2+x_2^2)(y_1^2+y_2^2)}+\sqrt{(x_3^2+\cdots+x_n^2)(y_3^2+\cdots+y_n^2)} \big)^2.
		\end{split}
	\end{equation}
	Due to randomness of data stored, we consider
	\begin{equation}\label{e:Kk3}
		\sigma(K_k)=(|\sigma(\vec{x}_k)||\sigma(\vec{y}_k)|+|\sigma(\vec{x}^c_k)||\sigma(\vec{y}^c_k)|)^2.
	\end{equation}
	Let $T_3=\{\vec{t}=(t_1,\cdots,t_{d^2})|\sum_{k=1}^{d^2} t_k=1, t_k\geq 0\}$. For $\vec{t}\in T_3$, we define the convex combination
\begin{equation}\label{e:Kt}
		K^{\vec{t}}=\sum_{k=1}^{d^2}t_kK_k.
	\end{equation}
	%the convex combination of $K_1,K_2,\cdots,K_{d^2}$.
%For example,
%	\begin{equation}\label{e:K19}
%K^{(0.1,0,0,0.9)}=0.1K_1+0.9K_4.
%	\end{equation}
	
	\emph{Theorem 3.}\label{t:K} For arbitrary observables $A$ and $B$, and a state $\rho$ on a $d$-dimensional Hilbert space $H$, let $\tilde{K}_k=\max_{\sigma\in\mathfrak{S}_{d^2}} \{\sigma(K_k)\}$ and $\tilde{K}=\max_{k}\{\tilde{K_k}\}$ for $k\in \{1,\cdots,d^2\}$. Then, we can obtain the following inequality:
	\begin{equation}\label{e:boundK2}
		\begin{split}
			(i)~ I^{c^{WYD}}_{\rho}(A)I^{c^{WYD}}_{\rho}(B)&\geq\tilde{K}\geq\tilde{K}_k\geq K_k\geq|\mathrm{Corr}^{c^{WYD}}_{\rho}(A,B)|^2;\\
			(ii)~	I^{c^{WYD}}_{\rho}(A)I^{c^{WYD}}_{\rho}(B)&\geq \max_{\vec{t}\in T_3}\{K^{\vec{t}}\} \geq|\mathrm{Corr}^{c^{WYD}}_{\rho}(A,B)|^2.\ \ \ \ \ \
		\end{split}
	\end{equation}
\begin{proof} It follows from the Cauchy-Schwarz inequality that
	%By the definition of $\tilde{K}_k$ and $\tilde{K}$, there is
	%	$ \tilde{K}\geq \tilde{K}_k\geq K_k $. Moreover,
	\begin{equation}
			\begin{split}
				K_k& = |\vec{x}_k|^2\cdot|\vec{y}_k|^2+|\vec{x}^c_k|^2\cdot|\vec{y}^c_k|^2 +2|\vec{x}_k|\cdot|\vec{y}_k|\cdot|\vec{x}^c_k|\cdot|\vec{y}^c_k| \\
				& \geq (\vec{x}_k,\vec{y}_k)^2+(\vec{x}^c_k,\vec{y}^c_k)^2 +2(\vec{x}_k,\vec{y}_k)(\vec{x}^c_k,\vec{y}^c_k)= (\vec{x},\vec{y})^2,
     \end{split}
    \end{equation}
where the equality holds if and only if $\vec{x}_k$ and $\vec{y}_k$ are proportional.
Also by the fundamental equality
    \begin{equation}
	\begin{split}
				K_k&\leq |\vec{x}_k|^2\cdot|\vec{y}_k|^2+|\vec{x}^c_k|^2\cdot|\vec{y}^c_k|^2 +(|\vec{x}_k|\cdot|\vec{y}^c_k| )^2+ (|\vec{y}_k|\cdot|\vec{x}^c_k|)^2=(\vec{x},\vec{x})(\vec{y},\vec{y}),
	\end{split}
	\end{equation}
where the equality holds if and only if $|\vec{x}_k||\vec{y}_k^{c}|=|\vec{x}_k^{c}||\vec{y}_k|$. Therefore,% Then we have
		\begin{equation}\label{e:Kk2}
			I^{c^{WYD}}_{\rho}(A)I^{c^{WYD}}_{\rho}(B)=(\vec{x},\vec{x})\cdot(\vec{y},\vec{y})\geq K_k\geq  (\vec{x},\vec{y})^2= |\mathrm{Corr}^{c^{WYD}}_{\rho}(A,B)|^2.
		\end{equation}
		Similarly, we also have
		\begin{equation}\label{e:Kk4}
			I^{c^{WYD}}_{\rho}(A)I^{c^{WYD}}_{\rho}(B)\geq \sigma(K_k)\geq |\mathrm{Corr}^{c^{WYD}}_{\rho}(A,B)|^2.
		\end{equation}
		This proves (a), from which (b) follows easily in view of convex functions.
\end{proof}
	
	Note that $\tilde{K}_k=\tilde{K}_{n-k}$ by \eqref{e:Kk3}, so it is enough to consider $\tilde{K}_k$ for $k=1, \cdots, [\frac n2]$.
	Set $\tilde{K}^q_k=q\tilde{K}_k+(1-q)\tilde{K}_n$ for $q\in [0,1]$, then we have
	\begin{equation}
		I^{c^{WYD}}_{\rho}(A)I^{c^{WYD}}_{\rho}(B)\geq\tilde{K}^q_k\geq\tilde{K}_k\geq|\mathrm{Corr}^{c^{WYD}}_{\rho}(A,B)|^2.
	\end{equation}
	
The arguments presented above demonstrate that our new uncertainty bounds, formulated in terms of convex functions, are tighter than previous ones. This is illustrated clearly in the following examples. We note that our uncertainty relations are state-dependent and apply to specific states in the quantum system. However, the comparison provided below is valid in general, and we have selected the state and observables solely for the purpose of illustration.
	
	\medskip

	\emph{Example 1.} Let $\rho:=\rho(\theta )=\frac{1}{2}(I+\vec{r}\cdot \vec{\sigma})$ be a mixed state with $\vec{r}=(\frac{\sqrt{3}}{3}\cos\theta, \frac{\sqrt{3}}{3}\sin\theta,0)$, where $\vec{\sigma}=(\sigma_x,\sigma_y,\sigma_z)$ is the vector of Pauli matrices. Consider observables
	$A=\sigma_x-\frac{1}{2}\sigma_z$ and $B=\sigma_x+\sigma_y+\sigma_z$.
	
\begin{figure}[!ht]
		%\centering
		\includegraphics[width=14cm,height=5cm]{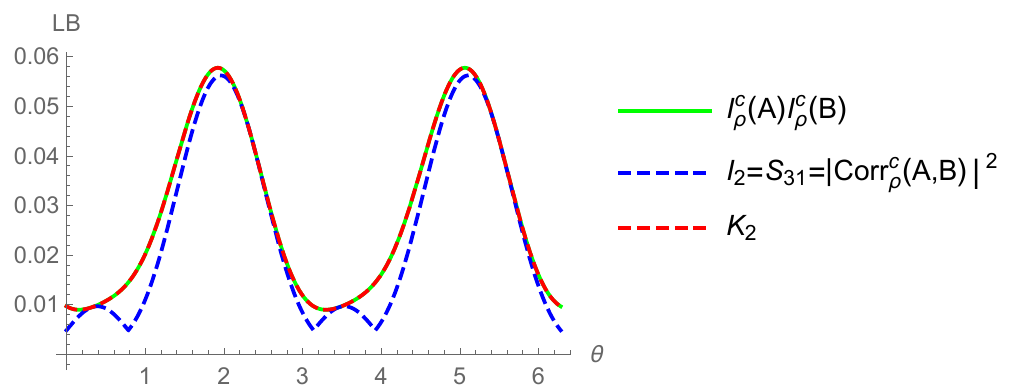}
		\caption{\footnotesize{\textbf{The lower bounds (LB) of $I^c_{\rho} (A)I^c_{\rho}(B)$ for the mixed state $\rho(\theta)$ in Example 1.} 
 $I^c_{\rho} (A)I^c_{\rho}(B)$, $|\mathrm{Corr}^c_{\rho}(A,B)|^2=I_2=S_{31}$ and our bound $K_2$ are shown in bold green, dashed blue
and red curves respectively.}}\label{gra3}
	\end{figure}

	Under the standard orthogonal basis $\{e_1=E_{11},e_2=E_{12},e_3=E_{21},e_4=E_{22}\}$, we have $\Gamma_{ij}=\mathrm{Corr}^{c}_{\rho}(e_i,e_j)=-\frac{1}{2}\Tr ([\rho^{\frac{1}{4}},e_i^{T}] [\rho^{\frac{3}{4}},e_j])$, where $c=c^{WYD}$ with $p=1/4$ (see \eqref{e:c}). The lower bounds $I_k$, $S_{pq}$ and $K_2$ can be calculated by \eqref{e:Ik}, \eqref{e:Spq} and \eqref{e:K2} respectively. Explicitly $I^c_{\rho} (A)I^c_{\rho}(B)=I_1=S_{41}=K_2$ and $|\mathrm{Corr}^{c}_{\rho}(A,B)|^2=I_2=I_3=I_4=S_{21}=S_{31}=S_{32}=S_{42}=S_{43}$. The bounds are drawn in figure \ref{gra3} which shows that
our bound $K_2$ is the best.
	
	To further demonstrate that our bounds are sharper than the previously available results, we chart the differences of our bounds with the previous bounds in the following example.
	\medskip
	
\begin{figure}[!ht]
		%\centering
		\includegraphics[width=14cm,height=6cm]{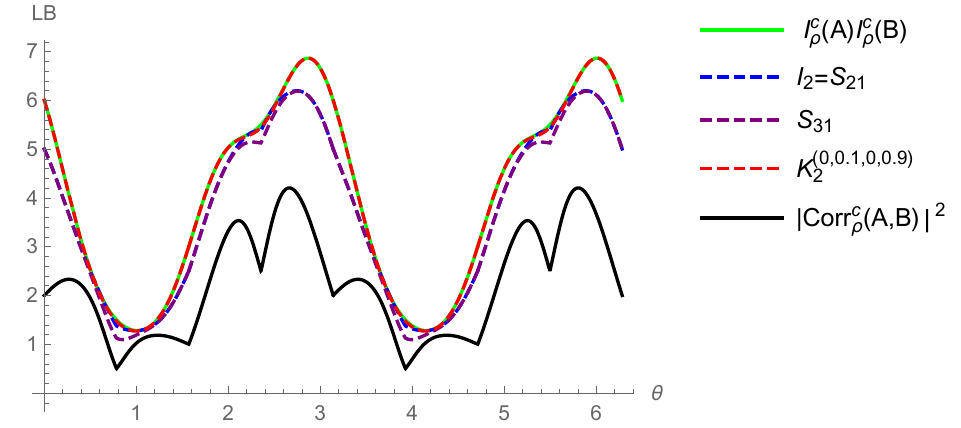}
		\caption{\textbf{The lower bounds of  $I^c_{\rho}(A)I^c_{\rho}(B)$ for the quantum state $|\psi(\theta)\ra$ in Example 2.}  
	 $I^c_{\rho}(A)I^c_{\rho}(B)$ and $|\mathrm{Corr}^c_{\rho}(A,B)|^2$
are shown in green and black bold curves respectively. The bounds $I_2$, $S_{31}$ and $K^{(0,0.1,0,0.9)}$
are shown in blue, purple and red dashed curves respectively, and are seen to be the closest to
the product.}\label{gra5}.
\end{figure}

\begin{figure}[!ht]
\includegraphics[width=8.5cm,height=5.5cm]{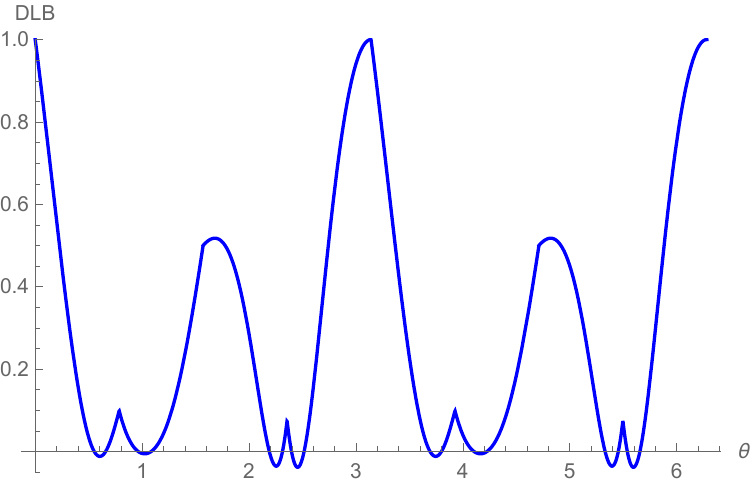}
\includegraphics[width=8.5cm,height=5.5cm]{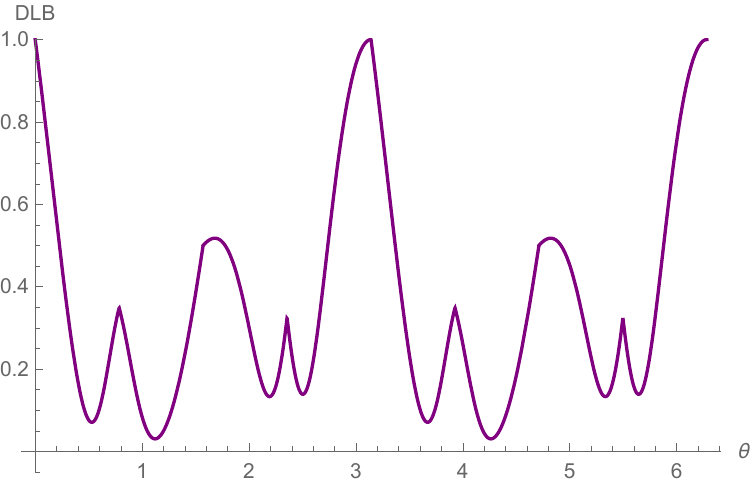}
		$ K_2^{(0,0.1,0,0.9)}-I_2 $\qquad\qquad\qquad\qquad\qquad\qquad\qquad\qquad\qquad\qquad
		$K_2^{(0,0.1,0,0.9)}-S_{31}$
		\caption{\textbf{The differences of the lower bounds (DLB) in FIG.4.} The blue and purple curves represent the differences $K_2^{(0,0.1,0,0.9)}-I_2$ and $K_2^{(0,0.1,0,0.9)}-S_{21}$ respectively, thus $K_2^{(0,0.1,0,0.9)}$ is the best bound.}\label{gra6}
	\end{figure}

	\emph{Example} 2.
	Consider the pure state $|\psi(\theta )\rangle=\cos\theta|0\ra-\sin\theta|2\ra$ on a 3-dimensional Hilbert space. Let observables $A$ and $B$ be
	\begin{equation*}
		\begin{split}
			A=\begin{bmatrix} 1&1-i&0\\1+i&-1&i\\0&-i&0\end{bmatrix}, \qquad 
			B=\begin{bmatrix}0&i&1-i\\-i&0&1\\1+i&1&0 \end{bmatrix}.
		\end{split}
	\end{equation*}
	The bounds $I_2$, $S_{31}$ and $K^{(0,0.1,0,0.9)}$ for the uncertainty relations (for $p=1/3$) are computed by (\ref{e:I2}), (\ref{e:S21}) and  (\ref{e:K19}), and
are shown in Figure \ref{gra5}. Their differences from the other bounds
are drawn in Figure \ref{gra6}, which clearly shows that ours are the strongest.
	
	%%%%%%%%%%%%%%%%%%%%%%%%%%%%%%%%%%%%
	\section{Improved uncertainty relations for sum-form skew information}
	In this section, we consider $m$ observables $\{A^i\}_{i=1}^m$ and a quantum state $\rho$ on a $d$-dimensional Hilbert space $H$.  Let $n=d^2$ as before. For each observable $A^i$, set $\vec{\a}^i= C_{\rho}{\vec{A}^i}=(\a_{i1},\cdots,\a_{in})^{\dag}$ and $\vec{X}^i=(x_{i1},\cdots,x_{in})^T$ with $x_{ij}=|\a_{ij}|$, where  $C_{\rho}$ is the matrix of the WYD-metric with respect to the state $\rho$ given as in \eqref{e:C1}. Then,
 \begin{equation}\label{sum}
   \sum_{i=1}^m I^{c^{WYD}}_{\rho}(A^i)=\sum_{i=1}^m|\vec{X}_{i}|^2=\sum_{i=1}^m\sum_{k=1}^n x_{ik}^2.
 \end{equation}
 Similarly, the elements $x_{ij}^2 (i=1,\cdots,m, j=1,\cdots,n)$ can be stored in a matrix. Let $M_{mn}=[x_{ij}^2]_{m\times n}$, and $B_2((a,b),(c,d))=|M_{mn}|_{\infty}-(x_{ab}^2+x_{cd}^2)+2x_{ab}x_{cd}$. Set $B_2=\max\{B_2((a,b),(c,d))|a,c\in\{1,\cdots,m\},b,d\in\{1,\cdots,n\}  \ \hbox{and}\ (a,b)\neq(c,d) \}$.
	For $q\in [0,1]$, we define
$\label{e:B2q}		B_2^q=qB_2+(1-q)\sum_{i=1}^m|\vec{X}^i|^2.$
Then we have the following strong uncertainty relation for the sum form WYD-metric adjusted skew information.

\emph{Theorem 4.} For arbitrary $m$ observables $A^i~ (i=1, \cdots, m)$ and a quantum state $\rho$ on
a $d$-dimensional Hilbert space $H$, then
	the following sum form uncertainty relation holds for the WYD-metric adjusted skew information:
	\begin{equation}
		\begin{split}
			\sum _{i=1}^mI^{c^{WYD}}_{\rho}(A^i)\geq B^q_2\geq B_2 \geq B_2((a,b),(c,d)),
		\end{split}
	\end{equation}
	
	\begin{figure}[!ht]
		\includegraphics[width=6cm,height=4cm]{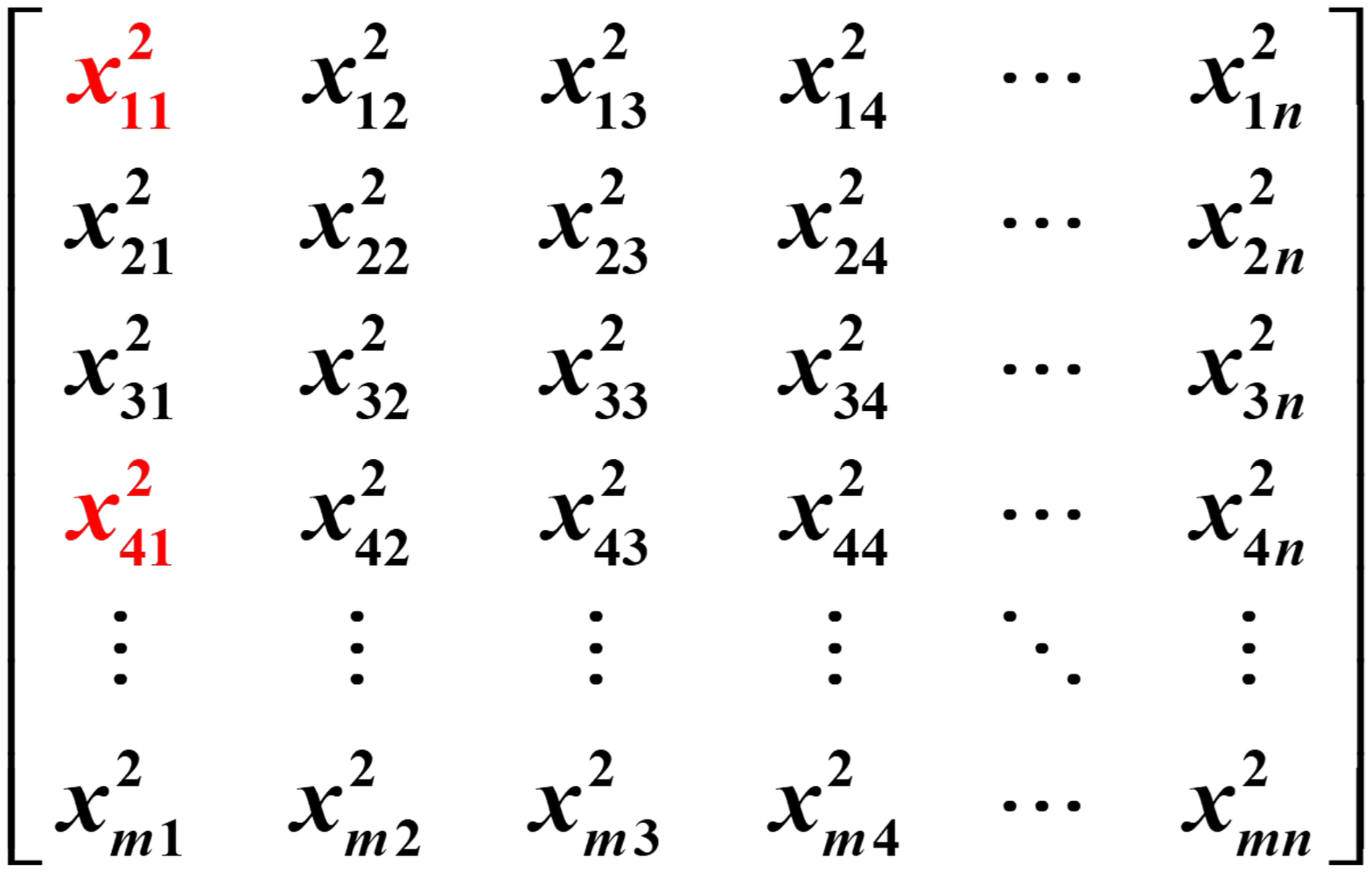}
		\caption{\textbf{Explanation of bound $B_2((a,b),(c,d))$.} 
%The bound $B_2((a,b),(c,d))=\sum_{i=1}^m\sum_{j=1}^n x^2_{ij}-(x_{ab}^2+x_{cd}^2)+2x_{ab}x_{cd}$, where  $i=1,\cdots,m, j=1,\cdots,n$.
The FIG.6 shows $B_2((1,1),(4,1))=\sum_{i=1}^m\sum_{j=1}^n x^2_{ij}-(x_{11}^2+x_{41}^2)+2x_{11}x_{41}$.}\label{gra7}
	\end{figure}
	
	\emph{Proof:} First it is clear that
	%By the fundamental inequality, %we have
	\begin{equation}\label{e:B22}
		\begin{split}
			B_2((a,b),(c,d))=\sum_{i=1}^m\sum_{j=1}^n x^2_{ij}-(x_{ab}^2+x_{cd}^2)+2x_{ab}x_{cd}
			\leq \sum_{i=1}^m\sum_{j=1}^n x^2_{ij} =\sum_{i=1}^mI^{c^{WYD}}_{\rho}(A^i).
		\end{split}
	\end{equation}
	\hskip\fill$\square$
	
	For $N$ real non-negative numbers $a^2_i ~(i=1,\cdots, N)$, according to the fundamental inequality,
	\begin{equation}
		\begin{split}
			N(\sum_{i=1}^N a_i^2)=\sum_{i=1}^N a_i^2+\sum_{1\leq i<j\leq N}(a_i^2+a_j^2)
			\geq (\sum_{i=1}^N a_i)^2, %\sum_{i=1}^n a_i^2 +\sum_{1\leq i<j\leq n}2a_ia_j
		\end{split}
	\end{equation}
	Therefore
	\begin{equation}
		\begin{split}
			\frac{m(m-1)}{2}\sum_{1\leq i<j\leq m}(|\vec{X}^i-\vec{X}^j|^2) \geq(\sum_{1\leq i<j\leq m}|\vec{X}^i-\vec{X}^j|)^2.
		\end{split}
	\end{equation}
	So we have the following uncertainty relation (cf. \cite{Ma})
	\begin{equation}\label{eq:M}
		\begin{split}
			\sum_{i=1}^m|\vec{X}^i|^2&=\frac{1}{2(m-1)}\sum_{1\leq i<j\leq m} (|\vec{X}^i+\vec{X}^j|^2+|\vec{X}^i-\vec{X}^j|^2)\\
			&\geq \frac{1}{2(m-1)}\big[\sum_{1\leq i<j\leq m} |\vec{X}^i+\vec{X}^j|^2+ \frac{2}{m(m-1)}(\sum_{1\leq i<j\leq m}|\vec{X}^i-\vec{X}^j|)^2\big].
		\end{split}
	\end{equation}
	For simplicity, set $L_{Ma}=\frac{1}{2(m-1)}[\sum_{1\leq i<j\leq m} |\vec{X}^i+\vec{X}^j|^2+ \frac{2}{m(m-1)}(\sum_{1\leq i<j\leq m}|\vec{X}^i-\vec{X}^j|)^2]$.
	
	\medskip
	
	\emph{Example} 3.
	Let us consider the mixed state $\rho:=\rho(\theta)=\frac{I_2+\vec{r}\cdot \vec{\sigma}}{2}$ with $\vec{r}=(\frac{3}{4}\sin\theta,0,\frac{3}{4}\cos\theta)$ and the following four observables $A^i ~(i=1,\cdots,4)$
	\begin{equation*}
		\begin{split}
			A^1=\begin{bmatrix} 1&2+i\\2-i&-1\end{bmatrix},\ \ \ \
			A^2=\begin{bmatrix} 1&i\\-i&-1 \end{bmatrix},
			A^3=\begin{bmatrix} 0&1+\frac{1}{2}i\\1-\frac{1}{2}i&0 \end{bmatrix},
   \ \ \ \
			A^4=\begin{bmatrix} 0&i\\-i&0 \end{bmatrix}.
		\end{split}
	\end{equation*}

\begin{figure}[!ht]
		\includegraphics[width=14cm,height=5.8cm]{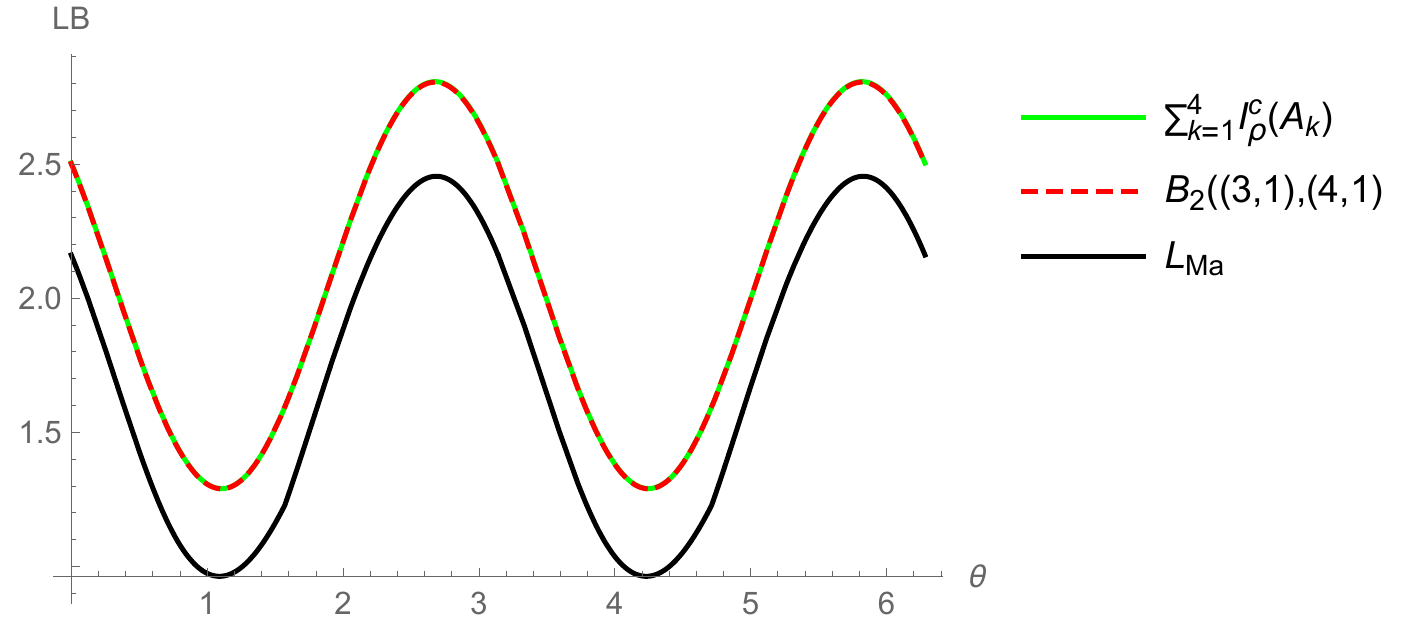}
		\caption{\textbf{The lower bounds of $\sum_{k=1}^4I_{\rho}^c(A^i)$ for the quantum state $\rho(\theta)$ in Example 3.} 
The quantity $\sum_{k=1}^4I_{\rho}^c(A^i)$, the bound $L_{Ma}$ from \cite{Ma} and our bound $B_2((3,1),(4,1))$ are in green, black and red-dashed curves respectively.}\label{gra71}
	\end{figure}

\begin{figure}[!ht]
		\includegraphics[width=13cm,height=5.2cm]{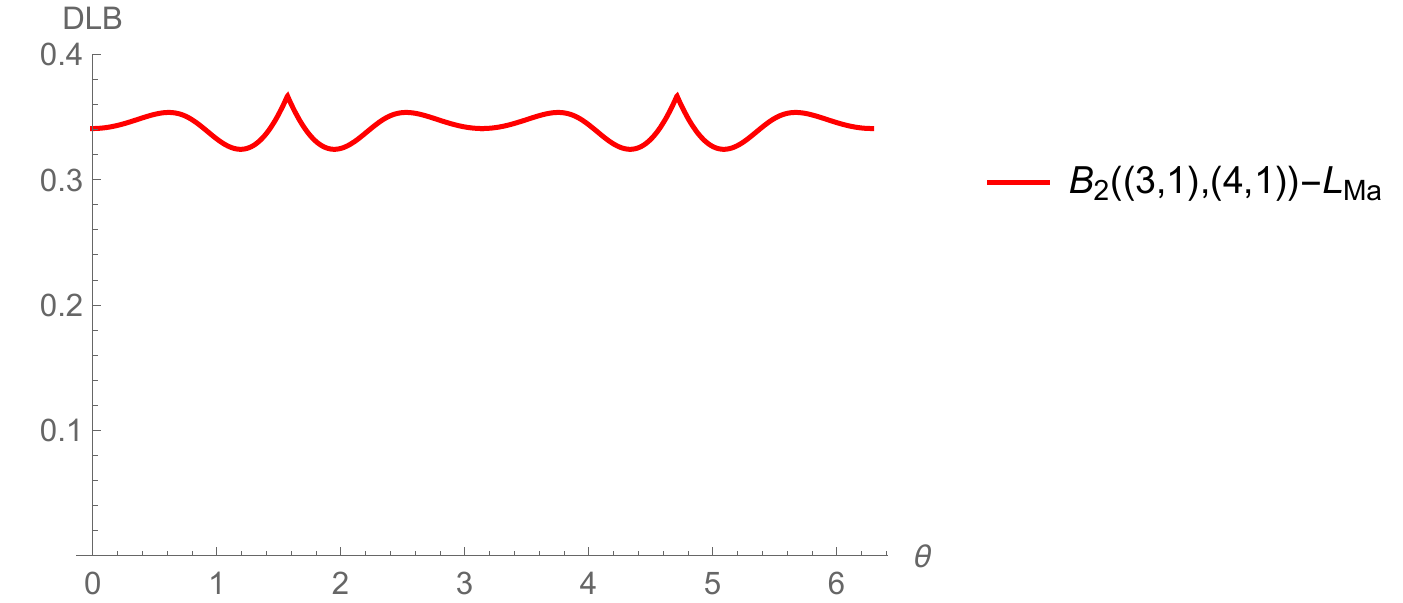}
		\caption{\footnotesize{\textbf{The difference of the lower bounds  in FIG.\ref{gra71}.}  The solid red curve represents $ B_{2}((3,1),(4,1))-L_{Ma}$, the difference of our new bound and the recent lower bound in \cite{Ma}, which reveals that our lower bound $B_{2}((3,1),(4,1))$ is tighter than that of $L_{Ma}$.}}\label{gra8}
	\end{figure}

	The bounds $B_2((3,1),(4,1))$ and $L_{Ma}$ can be calculated by (\ref{e:B22}) and \eqref{eq:M} respectively. They are drawn in FIG. \ref{gra71} and their differences are shown in FIG. \ref{gra8}, all are done with $p=1/3$ in defining the WYD-metric adjusted skew information. It happens that $B_2((3,1),(1,1))=\max\{B_2((a,b),(c,d))|a,c\in\{1,\cdots,m\},b,d\in\{1,\cdots,n\}  \ \hbox{and}\ (a,b)\neq(c,d) \}$ in this case. The graph shows that
our bound is tighter than the lower bound $L_{Ma}$ in \cite{Ma}.
	
	\medskip
	
	\emph{Example} 4. Let $|\psi(\theta)\rangle=\cos\theta|0\ra-\sin\theta|2\ra$ be a pure state on a 3-dimensional Hilbert space. Consider the four observables $A^i ~ (i=1,\cdots,4)$:
		\begin{equation*}
		\begin{split}
			A^1=\begin{bmatrix} 1&1-i&0\\1+i&-1&i\\0&-i&0\end{bmatrix},\ \ \ \
			A^2=\begin{bmatrix} 0&i&1-i\\-i&0&1\\1+i&1&0 \end{bmatrix}, 
			A^3=\begin{bmatrix} 0&0&1-i\\0&0&1\\1+i&1&0 \end{bmatrix},\ \ \ \
			A^4=\begin{bmatrix} 2&1-i&0\\1+i&-2&0\\0&0&0 \end{bmatrix}.
		\end{split}
	\end{equation*}

\begin{figure}[!ht]
		\includegraphics[width=14cm,height=5.6cm]{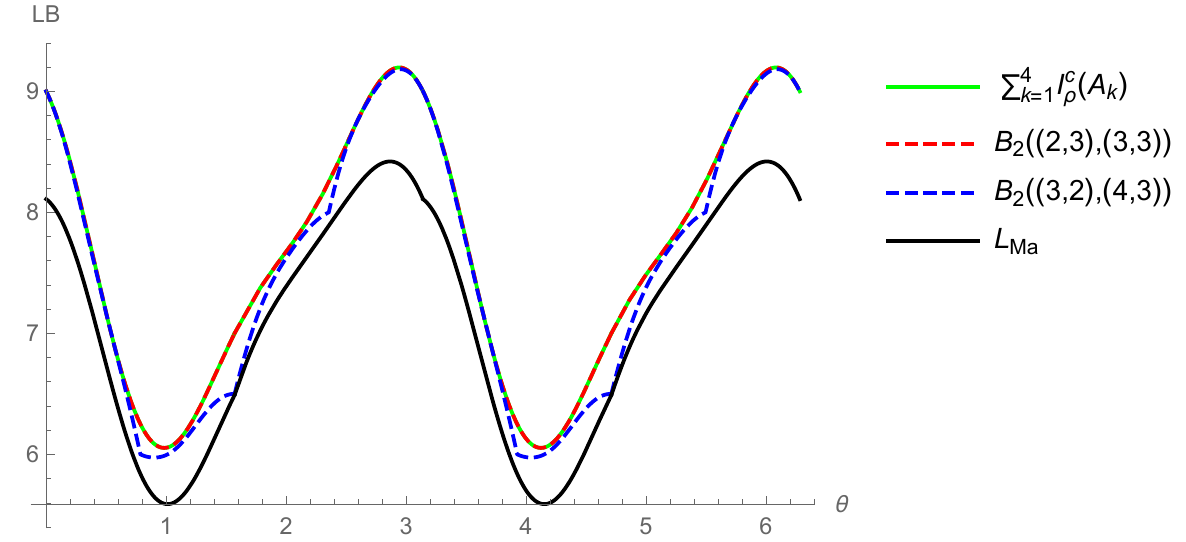}
		\caption{\footnotesize{\textbf{The lower bounds of $\sum_{i=1}^4 I^c_{\rho} (A^i)$ for quantum state $|\psi(\theta)\ra$ in Example 4.} 
The quantity $\sum_{i=1}^4 I^c_{\rho} (A^i)$, the bound $L_{Ma}$ from \cite{Ma}, our bounds $B_{2}((2,3),(3,3))$ and $B_{2}((3,2),(4,3))$ are shown in green, black, red (dashed) and blue (dashed) respectively.}}\label{gra9}
	\end{figure}

\begin{figure}[!ht]
\includegraphics[width=14cm,height=5.6cm]{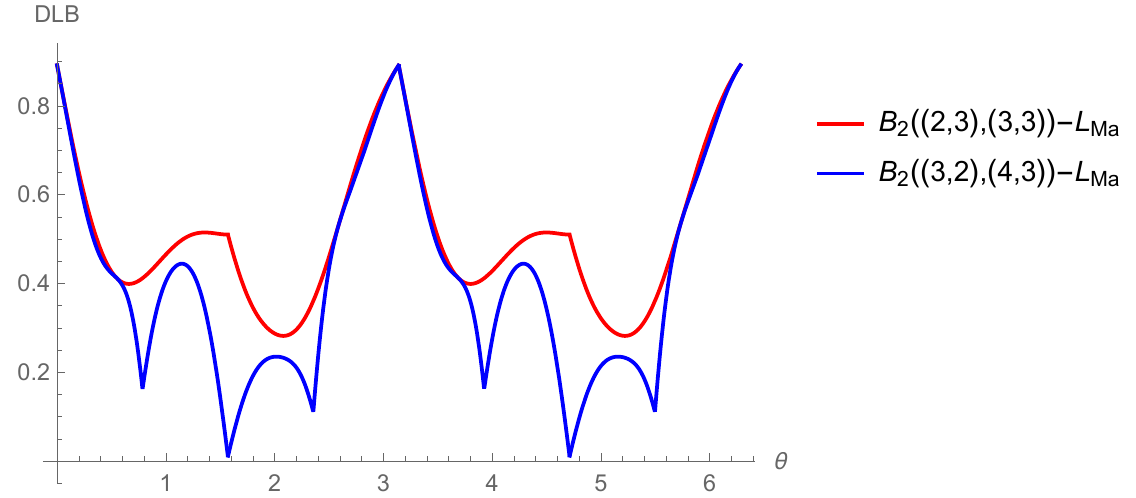}
\caption{\footnotesize{\textbf{The differences of the lower bounds in FIG.\ref{gra9}.} The solid red and blue curves are the differences $B_{2}((2,3),(3,3))-L_{Ma}$ and $B_{2}((3,2),(4,3))-L_{Ma}$, which shows that our bounds are tighter.}}\label{gra10}
	\end{figure}

	Similarly, the bounds $B_2((2,3),(3,3))$, $B_2((3,2),(4,3))$ and $L_{Ma}$ (with $p=1/3$) can be calculated by (\ref{e:B22}) and \eqref{eq:M}. $B_2((2,3),(3,3))=\max\{B_2((a,b),(c,d))|a,c\in\{1,\cdots,m\},b,d\in\{1,\cdots,n\}  \ \hbox{and}\ (a,b)\neq(c,d) \}$. The bounds are drawn in FIG. \ref{gra9} and the differences are drawn in FIG. \ref{gra10}. It is seen that our bound is tighter than the lower bound $L_{Ma}$ in \cite{Ma} and almost approximates the summation WYD-metric adjusted skew information.
	
	%%%%%%%%%%%%%%%%%%%%%%%%%%%%%%%%%%%%
	\section{\label{sec:leve4} Conclusion}
	
The Wigner-Yanase-Dyson-metric adjusted skew information was introduced by Hansen \cite{Hansen} as a measure of quantum information, based on the geometrical formulation of quantum statistics by Chentsov and Morozova, and earlier work by Wigner and Yanase. This quantity is non-negative and bounded by the variance of an observable in a state, and vanishes for observables that commute with the state.
In this work, we have reviewed and improved the ``fine-grained'' bounds for the product form of the Wigner-Yanase-Dyson-metric adjusted skew information, using sampling observable coordinates and appropriate convex functions. Additionally, we have proposed uncertainty relations for the sum form of $N$ quantum observables. Our method relies on straightening the key mathematical inequality underlying the uncertainty principle, resulting in new and stronger uncertainty relations in both the sum and product forms.
Our results are competitive with the recent sharp bound presented in \cite{Ma}, as demonstrated by several examples.

	\bigskip
	
	%%%%%%%%%%%%%%%%%%%%%%%%%%%%%%%%%%%%
	\centerline{\bf Acknowledgments}
	%The author is grateful to reviewers' comments that improve the quality of the paper.
	We would like to thank Dr. Yunlong Xiao for helpful discussions.
	The research is supported in part by NSFC grants 12226321, 12126351, 12126314 and 11871325, and by Foundation of Jianghan University as well as Simons Foundation
	grant no. 523868.
	\vskip 0.1in
	
	\textbf{Data Availability Statement.} All data generated during the study are included in the article.
	
	\textbf{Conflict of Interest Statement.} The authors declare no competing interests.
	
	%%%%%%%%%%%%%%%%%%%%%%%%%%%%%%%%%%%%%%%%%%%%%%%%%%%%
\bibliographystyle{amsalpha}

\end{document}